\documentclass[twocolumn,aps,showpacs]{revtex4}
\usepackage{amsfonts}
\usepackage{epsfig}
\usepackage{dcolumn}
\usepackage{graphicx}
\begin{document}

\title{Core Lexicon and Contagious Words}
\author{Dmitri Volchenkov and Philippe Blanchard,}
\affiliation{BiBoS Research Center, Bielefeld University, Postfach
100131, D-33501, Bielefeld, Germany  }
\author{ Serge Sharoff}
\affiliation{Russian Research Institute for Artificial Intelligence, \\
P.O.Box 85, 125190, Moscow, Russia }

\date{\today}

\begin{abstract}
We present the new empirical parameter $f_c$, the most probable
usage frequency of a word in a language, computed via the
distribution of documents over frequency $x$ of the word. This
parameter allows for filtering the core lexicon of a language
from the content words, which tend to be extremely frequent in
some texts written in specific genres or by certain authors.
Distributions of documents over frequencies for such words
display long tails as $x>f_c$ representing a bunch of documents
in which such words are used in abundance. Collections of such
documents exhibit a percolation like phase transition as the
coarse grain of frequency $\Delta f$ (flattening out the strongly
irregular frequency data series) approaches the critical value
$f_c$.

\end{abstract}

\pacs{89.70.+c, 05.40.-a, 05.45.Tp, 01.20.+x}

\maketitle

Studies of lists of words arranged in terms of their frequencies
belong to the most important domains of quantitative linguistics
\cite{Zipf}. Detection of most frequent words constituting the
core lexicon of a language is important not only for foreign
language learners, but also for various practical applications,
including text compression (this was recognized as early as in
\cite{S}), speech recognition \cite{Lau}, information retrieval
\cite{SJ}, etc. It is easy to order words with respect to their
mean frequency of uses, which is typically measured as the number
of instances of a word normalized by the sample of one million
words (ipm, instances per million words); though the notion of a
word should cover all word forms, like \textit{goes, went, gone}
for \textit{go.}

However, the mean frequency is not a sufficient selection
criterion, because of the large relative dispersion of the word
frequencies which vary very much from one text to the next
especially in ample and diverse collections of documents. Some
words (like prepositions) occur in many texts with predictable
rates, others (like pronouns or mental verbs) are significantly
more frequent for certain writers or genres, while some are
"contagious": these words (such as proper names, technical terms,
abbreviations, etc.) appear in just a few documents, but when they
appear, they are often found in abundance \cite{CG}. The
variability of rates of words can be characterized in a variety
of ways, including the Poisson K-mixtures \cite{CG}. It can be
measured by the coefficient of variation (the standard deviation
divided by the mean), as in \cite{Kilgariff}. However, the
coefficient of variation as a measure of relative dispersion is
not very useful when the average frequency is close to zero, which
occurs quite often for the semantically loaded words. Another way
to measure the variability of rates for contagious words is to
compute the \textit{document frequency} (or \textit{inverse
document frequency}, \cite{SJ}) by counting the number of
documents the given word is mentioned in, the \textit{burstiness}
parameter that is the mean frequency, except that it ignores
documents with no instances of the word (see references in
\cite{CG}), etc. As is evident, each of these parameters does not
capture much of the heterogeneous structure of word rates series
and none of them provides any general approach to describe it.

In this paper, we approach the problem of detection of contagious
words and selection for the core lexicon of a language from a
probabilistic point of view. In accordance to it, the frequency
$x$ of a word (counted as the number of its instances per million
words observed in any document of a given language) is a random
positive variable distributed  with some (unknown) probability
density function $\rho(x)$. Strongly irregular frequency data
series can be flattened out by introducing the frequency coarse
grain $\Delta f.$  The statistics of any word $w$ can be
characterized by the number of documents $N_w(n,\Delta f)=N\left[
(n-1)\Delta f<x\leq n\Delta f\right]$ for which the rate of the
word $w$ drops into the $n-$th frequency interval for various
$\Delta f$.

We have found that for any coarse grain $\Delta f$ the
distributions  of documents $N_w(n,\Delta f)$ over $n$ are the
asymmetric curves having one maximum ${f_c}$ that is the most
probable frequency at which the word $w$ would appear in a
randomly chosen document written in the given language. The value
of $f_c$ is independent of genres, authors, and topics of
documents and is an intrinsic characteristic of the word in a
contemporary language. Obviously, $f_c$ varies as the language
evolves approaching zero as the word becomes obsolete. For the
frequencies $x$ close to the distribution maximum $f_c$,
distributions of documents $N_w$ over $n$ are bell shaped, but
have anomalous tails as $|x-f_c|$ is large enough. We can
estimate the most probable frequencies $f_c$ of words
independently by two methods: first, from the distributions of
documents $N_w(n,\Delta f)$ as $\Delta f\to 1$ and, second, from
the distributions of authors using the same word in their texts.
For any word, both methods gave identical values of $f_c$.

The most probable frequency $f_c$ helps to detect the content
words and to select words for the core lexicon of a language.
Common words which appear uniformly in most documents (like
prepositions, conjunctions, relational verbs, some size
adjectives, etc.)  have usually relatively high mean rates
$\bar{f}$ (ipm), and $f_c\leq
\bar{f}$. They obviously belong to the core lexicon of a
language.

The use of semantically loaded words depends essentially on
authors, genres, and topics. Despite they are not found in a bunch
of documents, their mean rates $\bar{f}$ are still very high
because of their excessive popularity in certain collections of
texts, but their most probable rates get down $f_c\ll
\bar{f}$ indicating the presence of long  tails in distributions
of documents $N_w$ over $n$ as $n>f_c/\Delta f.$ Eventually, for
contagious words found in abundance in just a few documents, the
distributions $N_w$ over $n$ have long tails, however their mean
rates $\bar{f}$ are very low since $x=0$ for almost all texts, and
$f_c\approx \bar{f}$ or even $f_c>\bar{f}.$
\begin{table*}
\caption{Empirical parameters measuring the variability of rates:
the mean frequency $\bar{f}$ (ipm),  the number of documents in
which the word is used, the standard deviation of frequencies
$\delta f$, the coefficient of variance $\delta f/\bar{f}$, the
most probable frequency $f_c$ (ipm), the power exponents
$\alpha_w$ and $\beta_w$. Statistics on the $4\cdot 10^7$ words,
1566 texts written in Russian from 1980 to 2002.}
\label{Tab1}
\begin{ruledtabular}
\begin{tabular}{lccccccccc}
Group& lemma &$\bar{f}$  & No. of texts
  & $\delta f$
 &$\delta f/\bar{f}$  &$f_c$ &$\alpha_w$
 &$\beta_w$ \\
\hline
Relational& imetj (\textit{to have}) &715.38 & 1347 &717.21&
1.00 & 640 &1.089  & 3.433 \\
verbs &bytj (\textit{to be}) & 10635.78  &  1555  &  4481.04 &
0.43 & 9390 & 1.108 & 5.438 \\ \hline Motion& idti (\textit{to
go}) &1029.18& 1422 &   880.59 &   0.86&
900 & 0.956 & 3.536 \\
verbs& ehatj (\textit{to ride, to travel}) &221.36   &914 &448.66
&2.03 &128 &0.869 &2.149 \\ \hline Perception &smotretj
(\textit{to look at}) &817.28   &1284
&930.40 &1.14 &540 &1.201 &2.574 \\
verbs& slishatj (\textit{to hear})  &306.46    &1081   &370.51
&1.21 &220 &1.361 &3.249 \\ \hline Size & bolshoy
(\textit{large}) &1602.30    &1487    &908.81
&0.57 &1600 &0.978 &5.006 \\
   adjectives  &malenjkii (\textit{small})&386.17     &1173    &482.61 &1.25 &300
&0.915 &2.828 \\
& visokii (\textit{high})  &307.33     &1176    &404.18
&1.32 &300 &1.116 &3.016 \\
& niskii (\textit{low})  &73.01    &602 &172.55 &2.36    &60
&0.824 &2.043 \\ \hline Prepositions& v (\textit{in}) &28450.99
&1566 &8625.48 &0.30 &25200&1.346 &11.392 \\ \hline Conjunctions&
i (\textit{and}) &35196.38 &1566 &9620.76 &0.27 &32000&0.947
&11.648 \\ \hline
Pronouns &on (\textit{he})&17804.82  &1554 &10537.47 &0.59 &10400&0.952 &5.078 \\
& ona (\textit{she}) &6651.45 &1530 &6118.02 &0.92 &3300 &0.900
&3.006 \\ \hline
Abstract& vremya (\textit{time}) &1830.26    &1489 &1167.38 &0.64 &1800 &1.148 &6.331 \\
nouns &spravedlivostj (\textit{justice}) &41.91 &368 &158.17
&3.77 &24   &0.591 &1.762 \\ \hline
References&
stol(\textit{table}) &512.40 &1147 &629.38     &1.23
&300&1.222 &2.889\\
to objects& dom (\textit{house}) &1030.96 &1351 &1088.64  &1.06    &750
&0.867 &3.242 \\ \hline
References& professor&179.23 &502 &939.73   &5.24 &40   &0.636 &1.423\\
to people& intelligentsia& 62.64 & 284 & 1334.78& 21.31& 18 &
0.973& 1.021
\\  \hline
Contagious& KGB & 28.87 &207& 151.82& 5.26 & 48 & 1.011 &1.034 \\
words & Internet& 23.86 & 133 & 161.61 & 6.77 & 30 & 0.990 & 1.072
\end{tabular}
\end{ruledtabular}
\end{table*}

Our study of word rates is based on the reference corpus of
Russian \cite{corpus}, which includes more than $4\cdot 10^7$
words in $1566$ texts, which are balanced in their coverage of
various genres: fiction, newspapers, various informative texts
originally written in Russian from 1980 to 2002.  Unlike earlier
corpora (of about 1 million words), reference corpora of this
size are close to saturation, namely, any collection of new
documents added to the corpus does not cause statistically
significant changes to the frequency and patterns of uses of its
words.

The corpus informs us about the list of 5000 words most frequently
used in modern Russian \cite{5000}.  The study of $f_c$ shows
that for about 60$\%$ of them their most probable frequencies is
$f_c\leq 50$ ipm, and just 2.51$\%$ have $f_c\geq 600$ ipm. Among
the words having typically very high most probable frequencies,
one can mention conjunctions and prepositions, pronouns and
relational verbs, some motion verbs, and size adjectives. The
proper nouns, acronyms, technical terms and other semantically
loaded words typically have comparably small values of $f_c$.

Document counts in the $n-$th frequency interval obviously
decrease with $n$ as $n>f_c/\Delta f$. We have found that it
decays exponentially with $n$ for rather small coarse grains
$\Delta f \ll f_c,$
\begin{equation}\label{1}
N_w(n,\Delta f)\propto \exp \left[- \frac{n{\ }\Delta
f-f_c}{\xi_w}
\right],
\quad  \Delta f \ll f_c,
\end{equation}
where $\xi_w$ plays the role of a "correlation length" of the
word $w$ and diverges as the scale $\Delta f$ approaches the
critical value $f_c$ as $\xi_w\propto \left|f_c-\Delta
f\right|^{-\alpha_w}$ with the positive index $\alpha_w$ which is
close to unity for the majority of words (see the data of Table
\ref{Tab1}). The parameter $\xi_w$ casts the characteristic excess
of the word frequency $x$ over $f_c$.
\begin{figure}[ht]
\begin{flushleft}
\begin{minipage}[ht]{.36\linewidth}
\epsfig{file=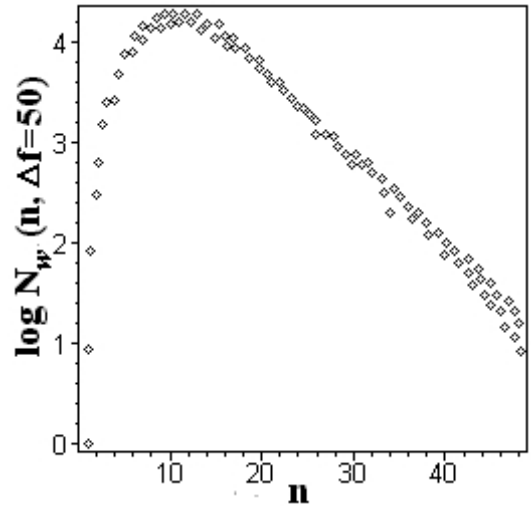, angle=0,width=7.0cm}
\end{minipage}
\end{flushleft}
\caption{\label{fig1} The distribution of documents over the frequencies
(instances per million words) for the relational verb
'imetj'(\textit{to have}). The most probable frequency for this
verb is $f_c=640$ instances per million words, the coarse grain is
taken as $\Delta f=50$. The distribution has an exponential tail.
Statistics on the $4\cdot 10^7$ words, 1566 texts written in
Russian from 1980 to 2002.}
\end{figure}

We have observed that the values of $\alpha_w$ for words belonging
to the same semantic group (such as relational verbs, motion
verbs, perception verbs, some size adjectives, pronouns,
conjunctions, and prepositions) are very close even if other
values of empirical parameters measuring the variability of their
rates are rather diverse.

For larger scales $\Delta f \approx f_c,$ for the frequency
intervals with $n>f_c/\Delta f$, the distributions $N_w$ are
scale free,
\begin{equation}\label{2}
N_w(n,\Delta f)\propto (n {\ }\Delta f-f_c)^{-\beta_w},
\end{equation}
where the index $\beta_w>1$ (see the data of Table \ref{Tab1}).
\begin{figure}[ht]
\begin{flushleft}
\begin{minipage}[ht]{.36\linewidth}
\epsfig{file=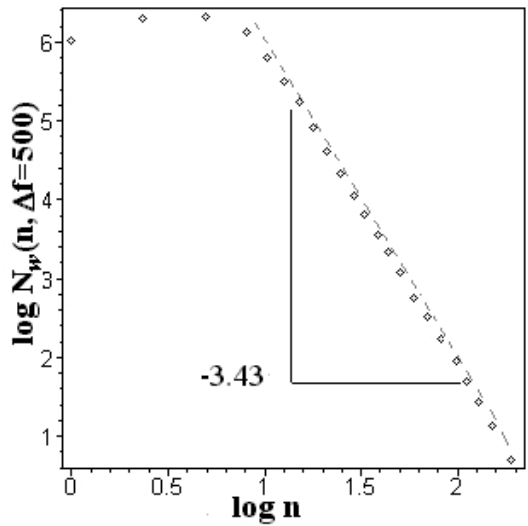, angle=0,width=7.0cm}
\end{minipage}
\end{flushleft}
\caption{\label{fig2} The distribution of documents over the frequencies
(instances per million words) for the relational verb
'imetj'(\textit{to have}) has a power law decaying tail with the
exponent $\beta=3.43$ when the coarse grain $\Delta f$ is taken
close to $f_c=640$.  Statistics on the $4\cdot 10^7$ words, 1566
texts written in Russian from 1980 to 2002.}
\end{figure}
The data of Table \ref{Tab1} show that the value of $\beta_w$
grows up with $f_c$ almost linearly (the coefficient of linear
correlation between $f_c$ and $\beta_w$ in Table \ref{Tab1} is
0.93). For the \textit{supercritical} phase $\Delta f\gg f_c,$
the tail of the probability distribution $N_w$ forms a stretched
exponential.

Let us note that the asymptotic behaviors (\ref{1}) and (\ref{2})
are typical for the \textit{subcritical} phase and the
\textit{critical} regime of percolation systems \cite{percol}.
Herewith, $\Delta f$ plays the role of the order parameter, and
$f_c$ is its critical value. A percolation-like phase transition
observed with respect to a word in the collections of documents
in which the frequency $x$ of this word exceeds its most probable
rate $f_c$ gives us an evidence of  existence of the genres of
literature. Content words are of particular interest since their
usage features usually the content of a text, so that the corpus
of texts in which such words pile up can be interpreted as the
literature genre possessing a special lexicon.

In this brief report we have studied the empirical distributions
of documents over the frequencies of Russian words computed on
the linguistic corpus of 1566 texts (of $4\cdot 10^7$ words). The
approach to the word frequency analysis which we have proposed is
very general and can be applied for any other human (or
artificial) language. We have introduced the new empirical
parameter $f_c,$ the most probable usage frequency of a word in a
contemporary language. The value of $f_c$ is independent on
authors, genres, and topics, but obviously varies in time as the
language evolves. The most probable frequency of a word could be
useful in studies devoted to the evolution of languages. This
parameter helps us to handle the heterogeneous structure of word
rate series and to determine whether they represent core lexicon.
Distributions of documents over frequencies for the semantically
loaded words which are found in abundance in a few documents have
remarkably long tails. The typical excess of the word rate over
$f_c$ which plays the role of the correlation length in (\ref{1})
can be used in the automatic recognition of grammatical functions
of words in a language. We have shown that the collections of
documents accumulating content words exhibit a percolation like
phase transition uprising the certain genres of literature
appropriate of specialized lexicons or terminologies.

One of the authors (S.S.) is supported by the Alexander von
Humboldt Foundation (Germany). D.V. benefits from the BiBoS
Research Center (Bielefeld University) as a guest researcher.
Authors thank T. Kr\"{u}ger for the fruitful discussion.


\begin{thebibliography}{000}

\bibitem{Zipf}
G. K. Zipf, \textit{Human behavior and the principle of least
effort. An introduction to human ecology}, Cambridge, MA:
Addison-Wesley (1949).


\bibitem{S}
C. E. Shannon, Bell. Syst. Tech. J. \textbf{27}, 379 (1948),
\textbf{27}, 623 (1948).

\bibitem{Lau}
R. Lau, R. Rosenfeld, S. Roukos, \textit{Adaptive Language
Modeling using the Maximum Entorpy Principle,} ARPA workshop on
Human Language Technology, Morgan Kaufmann Publ., San Francisco
(CA), ISBN 1-55860-324-7, 108 (1993).

\bibitem{SJ}
K. Spark Jones, J. of Documentation \textbf{28}, 11 (1972).

\bibitem{CG}
K. W. Church, W. A. Gale, {J. of Natural Language Engineering}
\textbf{1}, 2 (1995).

\bibitem{Kilgariff}
A. Kilgariff,
 International Journal of Lexicography \textbf{10}, 2 (1997).


\bibitem{corpus}
S. Sharoff, \textit{Proc. of the Corpus Linguistics Conference},
Lancaster (2001).

\bibitem{5000}
http://www.artint.ru/projects/frqlist/frqlist-en.asp

\bibitem{percol}
G. Grimmett, \textit{Percolation}, Springer Verlag, NY, Berlin,
Heidelberg (1989), D. Stauffer, A. Aharony, \textit{Introduction
to Percolation Theory}, Taylor $\&$ Francis, London, Washington
DC (1992).


\end{thebibliography}
\end{document}